\documentclass[preprint,showkeys,amssymb]{revtex4}
\usepackage{graphicx}

\begin{document}

  \title{Normal DGP in Varying Speed of Light Cosmology}

\author{A. Ravanpak}
\email{a.ravanpak@vru.ac.ir}
\affiliation{Department of Physics, Vali-e-Asr University, Rafsanjan, Iran}
\author{H. Farajollahi}
\email{hosseinf@guilan.ac.ir}
\affiliation{Department of Physics, University of Guilan, Rasht, Iran}
\affiliation{School of Physics, University of New South Wales, Sydney, NSW, 2052, Australia}
\author{G. F. Fadakar}
\email{gfadakar@guilan.ac.ir}
\affiliation{Department of Physics, University of Guilan, Rasht, Iran}

\date{\small {\today}}

\begin{abstract}
The varying speed of light (VSL) has been
used in cosmological models in which the physical constants vary over time. On the other hand, the Dvali, Gabadadze and Porrati (DGP) brane world model, especially its normal branch has been extensively discussed to justify the current cosmic acceleration. In this article we show that the normal branch of  DGP in VSL cosmology leads to a self-accelerating behavior and therefore can interpret cosmic acceleration. Applying statefinder diagnostics demonstrate that our result slightly deviates $\Lambda$CDM model.
\end{abstract}


\maketitle

\section{Introduction}
\label{sect:intro}

We know that any physical theory consist of at least one or more free parameters, called fundamental constants. The value of these parameters has been measured in observations and compared with theoretical predictions. Aside from some recent observational results which show the possibility of tiny variations of these constants, one can assume a varying constant theory and deals with its consequences \cite{Chand04},\cite{Bahcall04},\cite{Drinkwater98},\cite{Ubachs04},\cite{Petitjean04},\cite{Bertotti03}.

The varying constant theories have been proposed and studied in literature. For instance, Brans-Dicke (BD) gravity theory \cite{Brans61} which is an extension of standard general theory of relativity and considers a varying Newtonian constant $G$, by means of a scalar field. Barrow-Magueijo (BM) theory \cite{Barrow05} of varying electron-proton mass ratio $\mu\equiv m_e/m_p$, via a changing in electron mass using a scalar field. Bekenstein-Sandvik-Barrow-Magueijo (BSBM) scenario \cite{Bekenstein82},\cite{Sandvik02} where considers variations in the fine structure constant $\alpha$, driven again with a scalar field. Also, the one recently has been attracted a great deal of attention; the varying speed of light (VSL) theory, where as a cosmological model may be considered as a competitor to inflation, since it can solve some of the cosmological problems and provide a theory of structure formation. One can consider the VSL theory \cite{Moffat93},\cite{Magueijo03},\cite{Magueijo00},\cite{Barrow00} as a result of a varying-$\alpha$ theory, because of the relation between them, $\alpha=e^2/\hbar c$. If $\alpha$ varies, one of $e$, $\hbar$, $c$, or a combination of them has to be varied.

Although the constancy of the speed of light is the foundation of the theory of relativity and apparently it has been proved via many experiments such as Michelson-Morely experiment, one can still consider a VSL theory in the sense that the results of such experiments must still hold at the appropriate scale in this part of the Universe and at this time.

On the other hand, a large amount of recent studies investigate the effects of extra dimensions in our Universe \cite{Sami03},\cite{Farajollahi11},\cite{Nojiri00},\cite{Lopez10}. In the simplest model of higher dimensional gravity, called brane cosmology, we assume our four dimensional (4D) world as a brane embedded in a five dimensional (5D) spacetime \cite{Randall99a},\cite{Randall99b}. The DGP model is a especial case of brane cosmologies in which the 4D Universe is embedded in a 5D Minkowskian bulk \cite{Dvali00}. According to how one can embed the 4D brane into the 5D Minkowskian bulk, the DGP model includes two separate branches which are distinguished with a parameter $\epsilon=\pm1$. The case of $\epsilon=+1$, is dubbed self-accelerating branch, since it can show the late time acceleration without any dark energy component \cite{Deffayet01},\cite{Deffayet02}. But the case $\epsilon=-1$, called normal branch needs a dark energy component for late time acceleration. The most important feature of the DGP model is its self-accelerating branch which suffers from the ghost problem \cite{Nicolis04},\cite{Koyama06}. Thus, it will be very interesting if one can modify the normal branch in such a way that it becomes self-accelerating.
In \cite{Lopez09}, using a $f(R)$-brane in the DGP model the author has changed the normal branch to a self-accelerating one.

The effects of variation of physical constants in the context of different higher dimensional theories, have been investigated in recent years. In \cite{Brax03},\cite{Germani03}, respectively, varying constant theories in brane cosmology and in a string-inspired brane world model have been studied. Varying-$G$ scenario in brane cosmology is the main feature in \cite{Leon02},\cite{Amarilla10}. VSL in brane cosmology and in a brane-induced FRW Universe has been studied, correspondingly, in \cite{Youm01} and \cite{Alexander00}. Also, \cite{Steer02} studies VSL in brane scenario from a different point of view. But varying constant theories in the context of a DGP brane world model have not been investigated and the results and consequences of such a model are not clear yet.

In this manuscript we apply VSL scenario in the DGP brane world cosmology. Our aim is to study the effect of this modification on the normal branch of the DGP model to find out if the intergration of these two could lead the normal branch to be self-accelerating. The manuscript is outlined as follows. In Sec. II, we obtain our model equations in the presence of a varying-$c$. We should note that, the variation can be spatial or temporal or both. Here, we only discuss the variation with respect to time. In Sec. III, by assuming a widely used functionally for $c(t)$, we compare the normal DGP model in the presence of a constant $c$ and a time dependent $c(t)$. We constrain our model parameters under which the normal branch will be self-accelerating in a varying-$c$ theory. Sec. IV, includes conclusions and remarks.

\section{DGP VARYING SPEED OF LIGHT THEORY}

We start the DGP cosmologies within the framework of VSL theories with the metric
\begin{equation}\label{metric}
ds^2 = -n^2(t,y)c^2(t) dt^2+a^2(t,y)\gamma_{ij}dx^idx^j+b^2(t,y)dy^2
\end{equation}
where $\gamma_{ij}$ is the metric of a three dimensional maximally symmetric space with a constant curvature $k$, and $x^i$ are the coordinates on the spatial slices. The $a(t,y)$ is the cosmological scale factor on the brane and $b(t,y)$ can be considerred as the scale factor along the extra dimension. Also, we have assumed that the speed of light is only a function of time, $c(t)$.

Since in VSL theories the Lorentz invariance becomes clearly broken it is postulated that there exists a preferred Lorentz frame in which the action is similar to a usual Lorentz invariant action with a constant $c$, replaced by a field $c(x^\mu)$. It is called the principle of minimal coupling. In other words $c$, varies in the local Lorentzian frames associated with cosmological expansion. This effect is a special relativistic effect and not a gravitational one. So, as proposed in \cite{Albrecht99}, $c(t)$ does not introduce any corrections to the Einstein tensor for the above metric in this preferred frame and then we can derive the non-vanishing components of the 5D Einstein tensor as below
\begin{eqnarray}
  G_{00} &=& 3\left[\frac{1}{c^2(t)}\frac{\dot a^2}{a^2}-n^2\left(\frac{a'^2}{a^2}+\frac{a''}{a}\right)+k\frac{n^2}{a^2}\right] \\
  G_{ij} &=& \left[a^2\left(\frac{a'^2}{a^2}+2\frac{a''}{a}+\frac{n''}{n}+2\frac{a'n'}{na}\right)+\frac{a^2}{n^2c^2(t)}\left(-2\frac{\ddot a}{a}-\frac{a'^2}{a^2}+\frac{\dot a\dot n}{an}\right)-k\right]\gamma_{ij} \\
  G_{05} &=& \frac{3}{c(t)}\left(\frac{\dot an'}{an}-\frac{\dot a'}{a}\right) \\
  G_{55} &=& 3\left(\frac{a'^2}{a^2}+\frac{a'n'}{an}\right)-\frac{3}{n^2c^2(t)}\left(\frac{\ddot a}{a}+\frac{\dot a^2}{a^2}-\frac{\dot a\dot n}{an}\right)-3\frac{k}{a^2}
\end{eqnarray}
where dot and prime respectively mean derivative with respect to time $t$ and $y$.

In obtaining the above equations, we have assumed that the radius of the extra space is stabilized, i.e., $\dot{b} = 0$. Also, we have considered the $y$-coordinate is defined to be
proportional to the proper distance along the $y$-direction with $b$ being the constant of
proportionality, i.e., $b' = 0$. According to these assumptions, we have defined the $y$-coordinate such that $b = 1$.

By assuming all the matter fields are confined on the brane and using junction conditions, after some calculations we reach to
\begin{equation}\label{Friedmann1}
    H^2+\frac{kc^2(t)}{a^2(t)} = (\sqrt{\frac{8\pi G}{3}\rho+\frac{1}{4r_c^2}}+\frac{\epsilon}{2r_c})^2
\end{equation}
and
\begin{equation}\label{Friedmann2}
2\dot H +3H^2+\frac{kc(t)^2}{a^2}=-\frac{3H^2+\frac{3kc(t)^2}{a^2}-2\epsilon r_c\sqrt{3H^2+\frac{kc(t)^2}{a^2}}8\pi G}{1-2\epsilon r_c\sqrt{3H^2+\frac{kc(t)^2}{a^2}}}
\end{equation}
as the effective Friedmann equations on the 4D brane. Here, $\rho$ and $p$ are energy density and pressure of the matter fields, $G$ is the gravitational constant and $r_c$ is the crossover length scale which separates 4D and 5D regimes of the model.

The violation of energy conservation is a general feature of the VSL theory. It can be seen via combining the above two Friedmann equations
\begin{equation}\label{cons}
\dot\rho+3H\left(\rho+\frac{p}{c^2(t)}\right)=\frac{3kc(t)\dot c(t)}{4\pi Ga^2(t)}
\end{equation}
For $\dot c(t)\neq 0$, the conservation of energy is destroyed. So, any change in the speed of light may be considered as a source of matter creation. To solve this problem, the following two solutions have
been proposed. 1), We can modify the energy momentum $T_{\mu\nu}$ \cite{Shojaie06} by including other physical terms or
vary gravitational constant $G(t)$, such that
$G(t)c(t)^{-4} =const$ \cite{Barrow00}. Thus, the energy-momentum retains conserved. 2), we can neglect the
energy-momentum conservation, and regard the variation
of the speed of light as a source of matter creation \cite{Shojaie06}. In this paper, we adopt the latter and in the next section discuss the consequences.

\section{ The normal DGP branch in VSL}

Let us investigate the effect of VSL in the normal branch of DGP model. We start with the Friedmann equation of the normal branch in the original DGP, in which the speed if light $c$ is a constant,
\begin{equation}\label{f}
H^2+\frac{kc^2}{a^2(t)}=(\sqrt{\frac{8\pi G}{3}\rho+\frac{1}{4r_c^2}}-\frac{1}{2r_c})^2,
\end{equation}
where $\rho$ is the ordinary matter. Therefore, in the limit of late time, we can neglect it and the equation reduces to
\begin{equation}\label{f2}
H^2+\frac{kc^2}{a^2(t)}=0,
\end{equation}
or in terms of the new variable $\Omega_k=-k/(a^2H^2)$, as below
\begin{equation}\label{f3}
H^2=\frac{\Omega_{k0}H^2_0c^2}{a^2(t)}
\end{equation}
where the subscript, 'zero', represent the present value of parameters. Integrating this equation gives us the behavior of scale factor at late time as
\begin{equation}\label{a}
a(t)=(\sqrt{\Omega_{k0}}cH_0)t.
\end{equation}
Regardless of the values of $\Omega_{k0}$ and $H_0$, this relation shows no acceleration at late time.

Now, we do the same procedure in the presence of a varying $c(t)$. With attention to \ref{f}, we obtain at late time
\begin{equation}\label{fc}
H^2+\frac{kc^2(t)}{a^2(t)}=0,
\end{equation}
or
\begin{equation}\label{fc2}
H^2=\frac{\Omega_{k0}H^2_0c^2(t)}{a^2(t)}.
\end{equation}
In the following we assume the widely used expression for $c(t)$ as \cite{Barrow00}:
\begin{equation}\label{c}
c(t)=c_0a^n(t)=c_0(1+z)^{-n}
\end{equation}
where $c_0$, is the current value of the speed of light and
$n$, is a constant where for $n\rightarrow0$, approaches the constant speed of light limit. This is called the Machian scenario which has significant advantages to the phase transition scenario in which the speed of light varies abruptly at a critical temperature \cite{Moffat93},\cite{Albrecht99}. Also, since $\dot c/c=n\dot a/a$, the speed of light decreases in time for $n < 0$, and grows for $n > 0$. Replacing (\ref{c}) in (\ref{fc2}), one obtains
\begin{equation}\label{fc3}
    H^2=\frac{\Omega_{k0}H^2_0c_0^2a^{2n}(t)}{a^2(t)}.
\end{equation}
Integration leads to
\begin{equation}\label{a2}
a(t)=\left([\sqrt{\Omega_{k0}}c_0H_0(1-n)]t\right)^{\frac{1}{1-n}},
\end{equation}
where regardless of the values of $\Omega_{k0}$, $H_0$ and $c_0$,one can find the deceleration parameter as
\begin{equation}\label{q}
q=-\frac{\ddot a a}{\dot a^2}=-n.
\end{equation}

\begin{figure}
\centering
\includegraphics[width=0.50\textwidth]{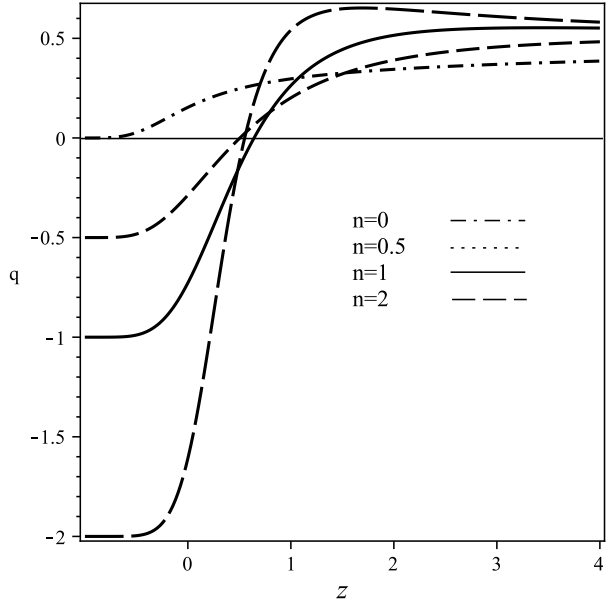}
\caption{The behavior of deceleration parameter versus redshift for different values of VSL-DGP parameter $n$. The case $n=0$, is related to an ordinary DGP model with constant speed of light. For $n>0$, the late time acceleration is obvious.}\label{fig2}
\end{figure}

According to \cite{Akarsu12}, the Universe would display power-law accelerating expansion for $-1<q<0$, exponential or De Sitter expansion for $q=-1$ and super-exponential expansion for $q<-1$. We know that our Universe is experiencing an accelerated expansion phase, so with attention to Eq.(\ref{q}), the normal DGP branch with a time varying speed of light as (\ref{c}), can naturally leads to late time acceleration for $n>0$. It approaches power-law, De Sitter and super-exponential acceleration for $0<n<1$, $n=1$ and $n>1$, respectively. The latter is related to the case when the Universe ends with big-rip \cite{Caldwell03}. The result of an ordinary normal DGP model with a constant speed of light is covered when $n=0$ (See Figure 1).

\section{statefinder diagnostic}

Statefinder diagnostic is an approach to distinguish different dark energy models. In this approach, two new geometrical variables related to the third derivative of scale factor with respect to time play the crucial role \cite{Sahni03}. In a non-flat Universe these variables are defined as
\begin{equation}\label{rs}
    r=\frac{\dot{\ddot a}}{aH^3}=q+2q^2-\frac{\dot q}{H}, \quad s=\frac{r-\Omega_t}{3(q-1/2)}\cdot,
\end{equation}
where $\Omega_t=1-\Omega_k $. We can rewrite above equation in terms of the equation of state parameter of dark energy, $w_d$, and its first time derivative as
\begin{equation}\label{rsw}
    r=\Omega_t+\frac{9}{2}w_d(1+w_d)\Omega_d-\frac{3}{2}\frac{\dot w_d}{H}\Omega_d, \quad s=1+w_d-\frac{1}{3}\frac{\dot w_d}{w_dH}.
\end{equation}

Thus, for the flat $\Lambda$CDM model, in which $w_d=-1$, we have $(r,s)=(1,0)$. As mentioned, the pair $(r,s)$, is usually used to discriminate different dark energy models. Also, one can compare the $(r,s)$ trajectories of these models with each other and study their deviation from $\Lambda$CDM model.

\begin{figure}
\centering
\includegraphics[width=0.45\textwidth]{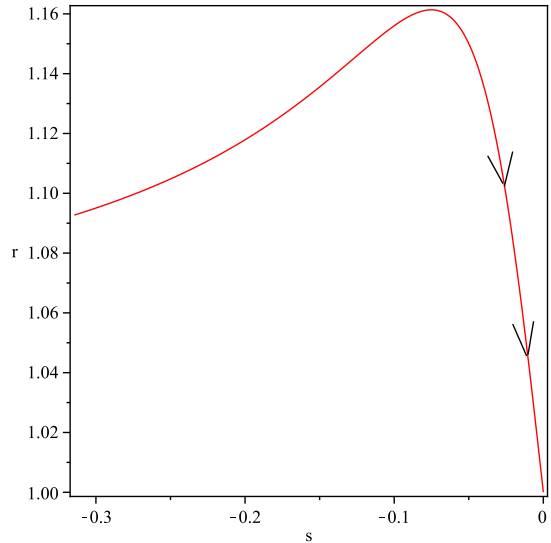}
\caption{The evolution of the statefinder parameter $r$ versus $s$, in non-flat VSL-DGP model with $n=1$. There is a very small deviation from the point $(1,0)$, related to $\Lambda$CDM model. This confirms analogue and closeness of the two models.}
\label{fig:rs}
\end{figure}

\begin{figure}
\centering
\includegraphics[width=0.45\textwidth]{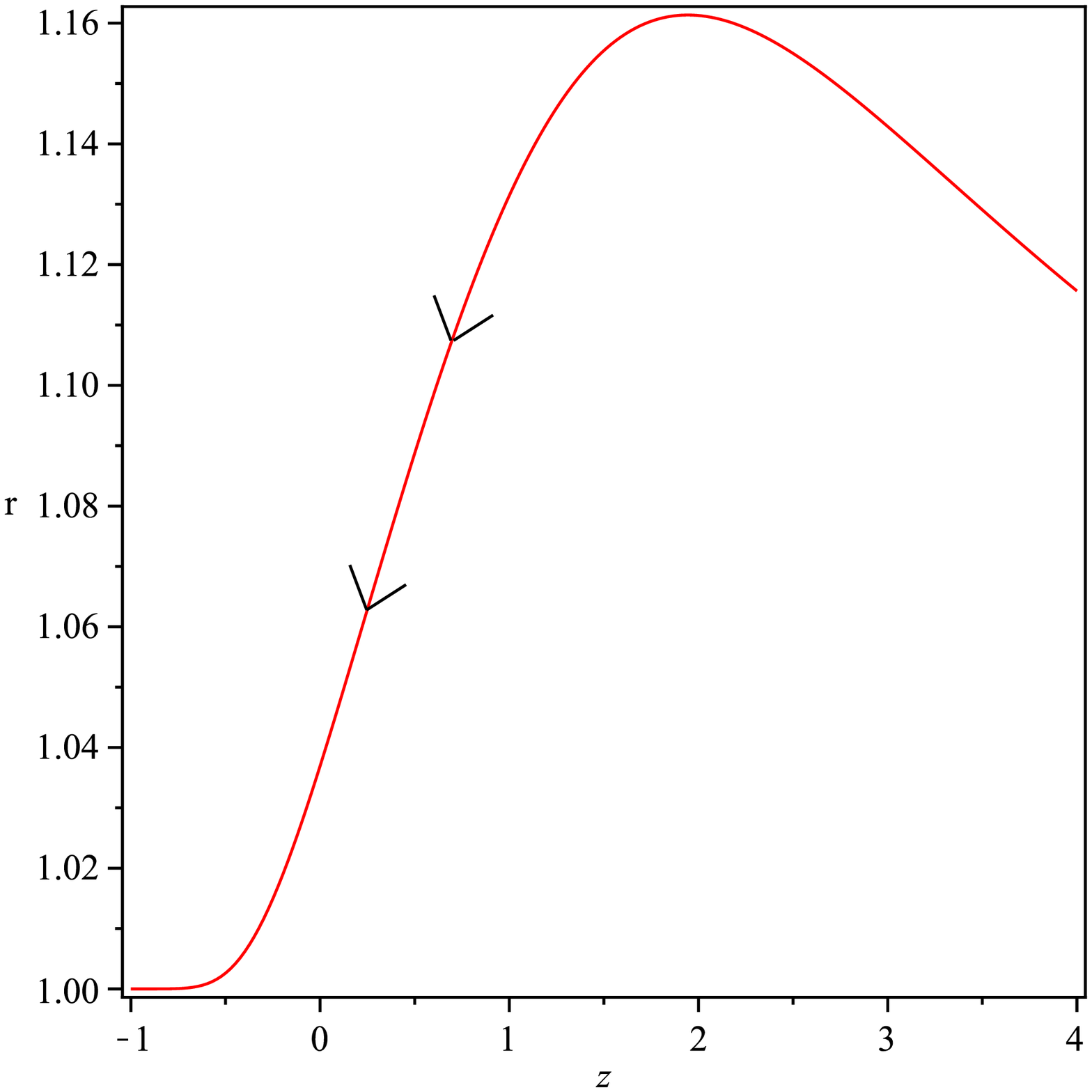}
\includegraphics[width=0.45\textwidth]{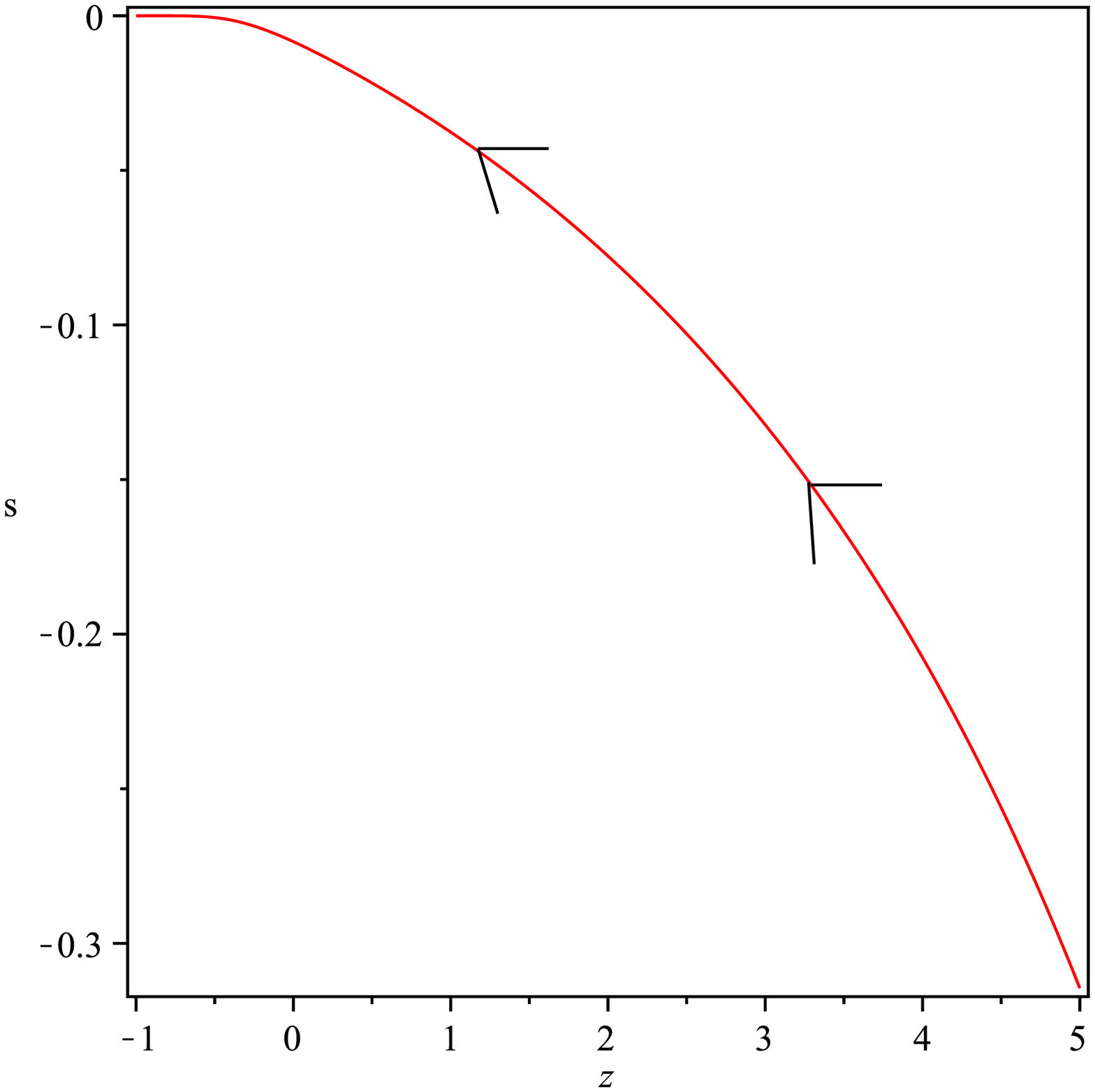}
\caption{The evolution of the statefinder parameters $r$ and $s$ versus redshift, in non-flat VSL-DGP model with $n=1$.}
\label{fig:rands}
\end{figure}

In our model, for the late time we have
\begin{equation}\label{r}
    r=-n+2n^2
\end{equation}
where we have used Eq.(\ref{q}). So we conclude that only for the case $n=1$, our model approaches the $\Lambda$CDM model and in a power-law acceleration $0<n<1$, or in a super-exponential acceleration $n>1$, the model deviates $\Lambda$CDM model.
Fig. \ref{fig:rs}, illustrates the trajectories belong to VSL-DGP model with $n=1$. The range of change of statefinder parameters, specially $r$, is  small, as it can be seen from Fig. \ref{fig:rands}, which means that our model has a tiny departure from $\Lambda$CDM model. Also, the curve $r(s)$, approaches the fixed point $(1,0)$ at late time.

\section{conclusion}

In this article we investigated the varying speed of light theory in the context of the normal branch of DGP brane cosmology. To this aim we considered a time dependent speed of light as $c_0a^n(t)$. We derived the modified Friedmann equations of the model. In comparison with the ordinary DGP model and in late time approximation we concluded that our model can experience a late time acceleration for $n>0$. We found that our model may lead to a power-law acceleration for $0<n<1$, an exponential acceleration for $n=1$ and also may end up with a big-rip for $n>1$. Using the statefinder diagnostic, we found that only the exponential or De Sitter expansion approaches the $\Lambda$CDM model.


\end{document}